\documentclass[runningheads]{llncs}
\usepackage[T1]{fontenc}
\usepackage{graphicx}
\usepackage{annotate-equations}
\usepackage{multirow}
\usepackage{amsmath}
\usepackage{xurl}
\usepackage{hyperref}

\begin{document}
\title{Characterising cardiac tissue properties with graph neural networks}
%
%
\author{Ching-En Chiu\inst{1}\orcidID{0009-0006-4419-0522} \and
Yoo Ri Kim\inst{2}\orcidID{0000-0003-3897-5869}\and
Magdi Saba\inst{3, 4}
\and
Danilo Mandic\inst{1}\orcidID{0000-0001-8432-3963}
\and
Marta Varela\inst{4, 5}\orcidID{0000-0003-4057-7851}
}
\authorrunning{C. Chiu et al.}
%
\institute{Department of Electrical and Electronic Engineering, Imperial College London, London, United Kingdom 
\and
Department of Internal Medicine, Chonnam National University College of Medicine, Gwangju, Republic of Korea
\and
St. George's University Hospital, London, United Kingdom
\and
Cardiovascular \& Genomics Research Institute, City St George's University of London, United Kingdom  \and
National Heart \& Lung Institute, Imperial College London, London, United Kingdom\\
\email{ching-en.chiu18@imperial.ac.uk}\\
\email{mvarela@citystgeorges.ac.uk}
}
\maketitle              
\begin{abstract}

Characterising electrophysiological properties of cardiac tissue efficiently and accurately from spatially sparse intracardiac measurements is clinically important for localising ablation targets and improving arrhythmia treatment. We developed a graph neural network-based framework trained on synthetic electrogram signals on 2D flat surfaces to identify areas of interest in the context of cardiac ablation for premature ventricular complexes (PVCs). Our method achieved an average precision of 0.96, 0.97, and 0.95 for the detection of single-patch fibrosis, rapid depolarisation and high excitability, respectively. The trained model can then be applied to 2D curved surfaces with few-shot fine-tuning, demonstrating its generalisation capability. Future work will develop this framework further for clinical use in PVC ablation. 

\end{abstract}
\section{Introduction}
Premature ventricular complexes (PVCs) are amongst the most common arrhythmias, detected in 40–75\% of apparently healthy individuals on ambulatory Holter monitoring and affecting approximately 1–4\% of the general population~\cite{ng2006treating}. They are characterised by early ventricular depolarisations that originate outside the normal conduction pathway. A small proportion of patients have symptoms severe enough to require treatment or a sufficiently high PVC burden to raise concerns about PVC-induced cardiomyopathies. 

Cardiac ablation, a common treatment option, involves creating a focal lesion that eliminates or electrically isolates the tissue responsible for the initiation or propagation of the abnormal electrical signals. The arrhythmogenic substrate is typically a small focal region of viable myocardium with abnormal electrophysiological properties (such as abnormal calcium handling or enhanced catecholamine sensitivity). In addition, PVCs in structural heart disease may arise at scar border zones, where surviving myocyte bundles, fibrosis, slow or heterogeneous conduction, and altered cell-to-cell coupling facilitate abnormal automaticity or re-entry.

During the ablation procedure, catheters equipped with electrodes are typically inserted into the heart to measure electrical signals at the endocardium, known as contact electrograms (EGMs). Current approaches to identify ablation targets from EGMs in a clinical setting remain largely empirical and rely heavily on the operator’s expertise. A common strategy for PVCs is \textit{activation mapping}, which identifies the earliest local activation site. The ablation outcome is often limited by infrequent intraprocedural PVCs, however. When PVCs are absent, \textit{pace mapping} may be adopted. It relies on the assumption that pacing at the PVC origin should produce waveforms similar to the clinical PVC~\cite{enriquez2024mapping}. Nevertheless, PVCs are often not inducible even with deliberately pacing, with~\cite{baser2014infrequent} reporting only 7\% of reproducibly inducible PVCs amongst patients with infrequent PVC during procedures. It is therefore useful to be able to locate ablation targets based on tissue properties during sinus rhythm. This would increase the procedure's success and reduce procedural time by limiting the need for extensive mapping.


\textbf{Aim} This study aims to characterise the underlying electrophysiological properties of the myocardium from spatially sparse EGM signals, using convolutional (CNNs) and graph neural networks (GNNs). Our framework identifies, at high spatial resolution, regions of fibrosis, abnormal repolarisation and enhanced automaticity. The proposed method may help improve ablation procedures for PVCs in the future.

CNNs excel at signal classification tasks, including electrocardiogram (ECG) analysis~\cite{li2017classification}, whereas GNNs are a specialised class of deep learning technique used on data represented as graphs. GNNs leverage \textit{message-passing} operations where each node iteratively exchanges information with its neighbours and update its own state. They have been used for cardiac electrophysiology applications, in which cardiac geometries are represented as graphs. For example, reconstructing atrial fibrillation dynamics from sparse contact mapping measurements~\cite{jenkins2025mapping} and modelling action potential propagation~\cite{morier2025learning}. The latter work particularly highlights GNNs' ability to generalise to complex geometries after only being trained on simple ellipsoid meshes, hinting at fast patient-specific cardiac applications.

70-80\% of idiopathic PVCs originate in the Right Ventricular Outflow Tract (RVOT)~\cite{calvo2013radiofrequency}. As RV is relatively thin, it is also well suited for the development and testing of methods that encode the geometry of myocardial tissue with triangular meshes, a type of graph.

Several automated approaches have been proposed to support PVC ablation. These methods can improve the reproducibility and efficiency of activation mapping, but they remain dependent on the presence of PVCs during the procedure and primarily aim to localise the ectopic origin without necessarily characterising the underlying electrophysiological properties. This is done either coarsely, using the surface ECG~\cite{wang2026advancements} or, at higher resolution, from intraprocedural EGMs. Examples of EGM-based automation include automated activation mapping, in which algorithms are developed to annotate local activation times using fixed features such as bipolar EGM onset or the maximal negative slope of the unipolar EGM~\cite{alcaine2018automatic,acosta2018clinical,jauregui2021manual}. In contrast, our work aims to identify pathological regions associated with abnormal electrophysiological properties, providing a potential route towards substrate-informed ablation planning when intraprocedural PVCs are infrequent.


\section{Methods}
All code used in this study is available at \url{https://github.com/annien094/GNN_cardiac_fibrosis.git}.

\subsubsection{Data Generation}
We use synthetic data to train and test the performance of our CNN and GNN-based tissue characterisation. For this, we solve the isotropic Aliev-Panfilov monodomain model~\cite{aliev1996simple} in 2D geometries representative of RV patches. We solve for the transmembrane potential, $V(\vec{x},t)$ and a latent field $W(\vec{x},t)$ that controls the recovery of the action potential:
\begin{subequations}\label{eq:AP eqs}
\begin{equation}\label{eq:AlievPanfilov1}
\frac{\partial V}{\partial t} = \nabla \cdot (\mathbf{D} \nabla V) - kV (V - a) (V - 1) - VW
\end{equation}
\begin{equation}\label{eq:AlievPanfilov2}
\frac{\partial W}{\partial t} = \left( \epsilon + \frac{\mu_1 W}{V + \mu_2} \right) \left( -W - kV (V - b - 1) \right).
\end{equation}
\end{subequations}

Here, $\mathbf{D}$ is the diffusion tensor which reduces to a scalar diffusion coefficient in isotropic and homogeneous conduction we use to model healthy tissue. $k$ controls the steepness of depolarisation, and $a$ is the excitation threshold potential. A no-flux Neumann condition: $\frac{\partial V}{\partial \vec{n}}=0$ is enforced at the boundary. The parameter values are chosen as in~\cite{goktepe2010electromechanics} to model the healthy tissue.

We solve the action potential in the following 2D geometries:
\begin{itemize}
    \item A square (dimensions: $100 \times 100 ~\text{mm}^2$), using a central finite differences solver ($\text{d}x = 1$ mm) and 4-step Runge-Kutta method ($\text{d}t=0.01~\text{TU}=0.129~\text{ms}$).
    \item Two curved 2D surfaces corresponding to ellipsoid patches of comparable physical size (surface area $\approx 420$ and $440~\text{mm}^2$) but different curvature (maximum Gaussian curvature $K = 1.19\times10^{-2}$ and $2.23\times10^{-3}~\text{mm}^{-2}$ for the higher- and lower-curvature cases, respectively). Mesh size: 19,481 nodes, 38,400 triangles, triangle size: $0.17 \pm 0.05$ mm). An implicit-explicit time-stepping scheme is implemented in FEniCS~\cite{baratta_2023_10447666}.
\end{itemize}


We pace every 645 ms (50 TU in the Aliev-Panfilov model) for 3 times, each at a random different location that is outside of any pathological regions. The total duration of each recorded signal is 1935 ms, with a sampling frequency of 775 Hz.

\subsubsection{Electrogram (EGM) Computation}
From the action potential field, $V (\vec{x}, t)$, we can compute the (unscaled) unipolar extracellular potentials, $\phi_e(\vec{x_e}, t)$, measured by electrodes placed on the endocardium at locations $\vec{x_e}$ through an unbounded Poisson equation solver~\cite{plonsey2007quantitative}:
\begin{equation}\label{eq:egm}
    \phi_e(\vec{x_e}, t) \propto \iint \frac{\nabla \cdot (\mathbf{D}\,\nabla V (\vec{x}, t))}
{|\vec{x} - \vec{x_e}|}
\, \text{d}S.
\end{equation}
In the square geometry, $\phi_e(\vec{x_e}, t)$ are calculated at 100 locations placed uniformly across the square. For the curved tissue, $\phi_e(\vec{x_e}, t)$ are calculated at 192 locations uniformly distributed across the surface, offset by a small distance ($0.01$ mm) from the tissue surface along the outward normal.

\subsubsection{Modelling of Focal Pathology}

We introduce heterogeneities in model parameters to model localised pathological changes that could be related to the sources or pathways of PVCs, including:
\begin{itemize}
    \item Fibrosis: assigned a lower diffusion coefficient $D$~\cite{roy2018image} sampled uniformly from [0.02, 0.04] $\text{mm}^2\,\text{TU}^{-1}$, 60-80\% lower than the baseline $D=0.1~\text{mm}^2 \,\text{TU}^{-1} $.   
    \item Fast depolarisation: assigned a higher $k$ sampled uniformly from [9.6, 16.0], 20-100\% higher than the baseline $k=8.0$.
    \item High excitability: assigned a negative excitation threshold $a=-0.025$. These regions thus have the ability to initiate action potentials, although they are not sources of action potentials in our model -- the frequency of spontaneous depolarisation at the chosen $a$ is lower than the external pacing frequency of 7.75 Hz.
\end{itemize}


For each type of abnormality, we generate 6000 simulations. Each simulation is independently generated with heterogeneities of different irregular polygonal shapes, location, dimensions, and, for fibrotic patches, a different number of heterogeneities. The shapes are generated by sampling $n_v \in [6, 12]$ vertices at random radii around a circle, applying optional elongation, and enforcing separation.



\subsubsection{Deep Learning Model for Focal Pathology Detection}
\begin{figure}[htp!]
    \centering
    \includegraphics[width=1.0\linewidth]{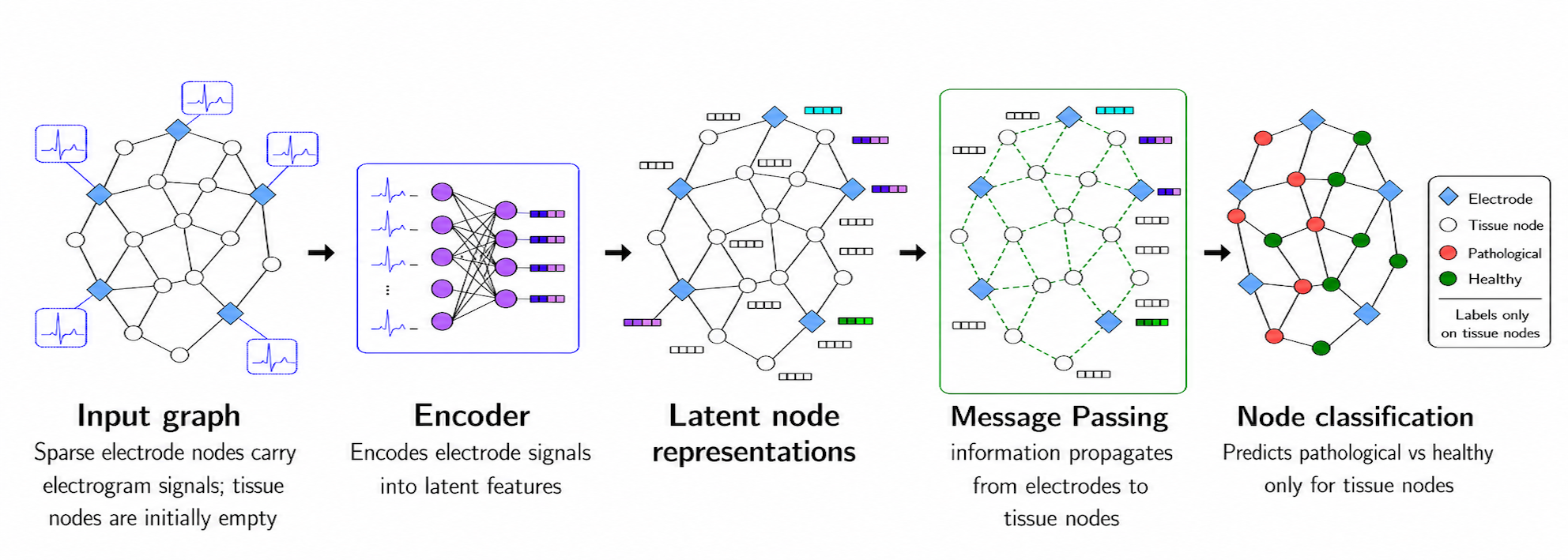}
    \caption{A schematic of our proposed framework. The temporal encoder module based on a CNN learns latent representations of the EGM waveforms. Then, the GNN based spatial module propagates the latent information from sparsely distributed electrodes to high-resolution tissue nodes. A small classification head then takes the final embeddings and performs binary classification.}
    \label{fig:model arc}
\end{figure}

To detect, from the EGM signals, the location of focal pathology introduced in the Aliev-Panfilov model, we use a deep learning model composed of two modules (Fig.~\ref{fig:model arc}):
\begin{enumerate}
    \item A temporal encoder module based on a CNN to learn low-dimensional latent representations of the EGM waveforms, and
    \item A spatial module based on a GNN to propagate the information in the sparsely distributed latent EGMs waveforms to a high-resolution representation of the myocardium.
\end{enumerate}

\paragraph{CNN Module}
The temporal module is composed of two
blocks, each with two 1D-CNN layers (kernel sizes 64 and 32, respectively), followed by max pooling, group norm, and dropouts, producing a 64-dimensional latent embedding for each EGM trace. Designed as an encoder to be coupled to the spatial module, it is initially pretrained on its own on a per-electrode binary EGM classification task: from a single electrode's $\phi_e$ time series, it predicts whether that electrode overlies an abnormal region (characterised by either lower $D$, higher $k$, or negative $a$).


\paragraph{GNN Graph Construction}
We represent the spatial relationships between each tissue element and the electrodes using a graph. We describe two types of nodes, though the model does not explicitly distinguish between them: 1) $i_\phi$, the electrodes, at which we have $\phi_e$ measurements, 2) $i_V$, the tissue nodes. Each $i_\phi$ node is connected it to its 20 nearest $i_V$ neighbours. Apart from the connections to $i_\phi$, each $i_V$ is connected to all the other $i_V$ nodes within a radius of $1.0$ mm.

For the 2D square, we have 3600 $i_V$ and 100 $i_\phi$ nodes ($i_V$ are $6 \times$ finer than the electrodes along each axis). For the 2D curved surfaces, we have 1{,}728 $i_V$ 192 $i_\phi$ nodes ($i_V$ are around 3$\times$ finer than the electrodes along each direction). The same graph resolution is used for both the higher- and lower-curvature geometries.


\paragraph{GNN Graph Attention Layers}
Each $i_\phi$ node is initialised with the 64-dimensional latent representation from the CNN Module. $i_V$ node initialisation is set by taking the bilinear interpolation of neighbouring $i_\phi$ initialisations. For information exchange across the graph nodes, we adopt two graph attention (GATv2) layers \cite{brody2021attentive}, which allows each node to learn to pay different level of attention to its neighbours during two rounds of message passing. The attention is learnt and initialised based on edge features, where the edge features are the inverse of distances between nodes, similar to the formulation in Eq.~\ref{eq:egm}. 



\paragraph{Experimental Setup}\label{training pipeline}
80\% of the simulations were used for training, leaving 20\% for testing. The loss function is a combined cross entropy and Dice score loss, equally weighted, calculated on the $i_V$ nodes. The CNN is trained using an Adam optimiser for 100 epochs, whereas the GNN is trained for 50, both with early stopping.


To assess the performance of the GNN module, we compare the performance of the full CNN+GNN (`High Resolution (HR) Model' below) approach against a CNN-only classification (`Low Resolution (LR) Model' below). For this comparison, after pretraining the temporal CNN module, the Low Resolution Model assigns each $i_V$ the label of its nearest $i_\phi$. We also evaluate both methods' robustness against noisy signals and electrode sparsity.


Finally, to investigate the GNN models' generalisation capability, we apply the model trained on 2D flat surfaces directly on simulations on curved surfaces to detect focal pathology, without any further training (`zero-shot'). Then, we fine-tune the model over 10 additional simulations on the lower-curvature surface, each with a different pathology location (`few-shot'), and apply it to detect focal pathology again on held-out graphs.

\section{Results}

\subsubsection{2D Square}

\begin{figure} [htp!]
    \centering
    \includegraphics[width=1.0\linewidth]{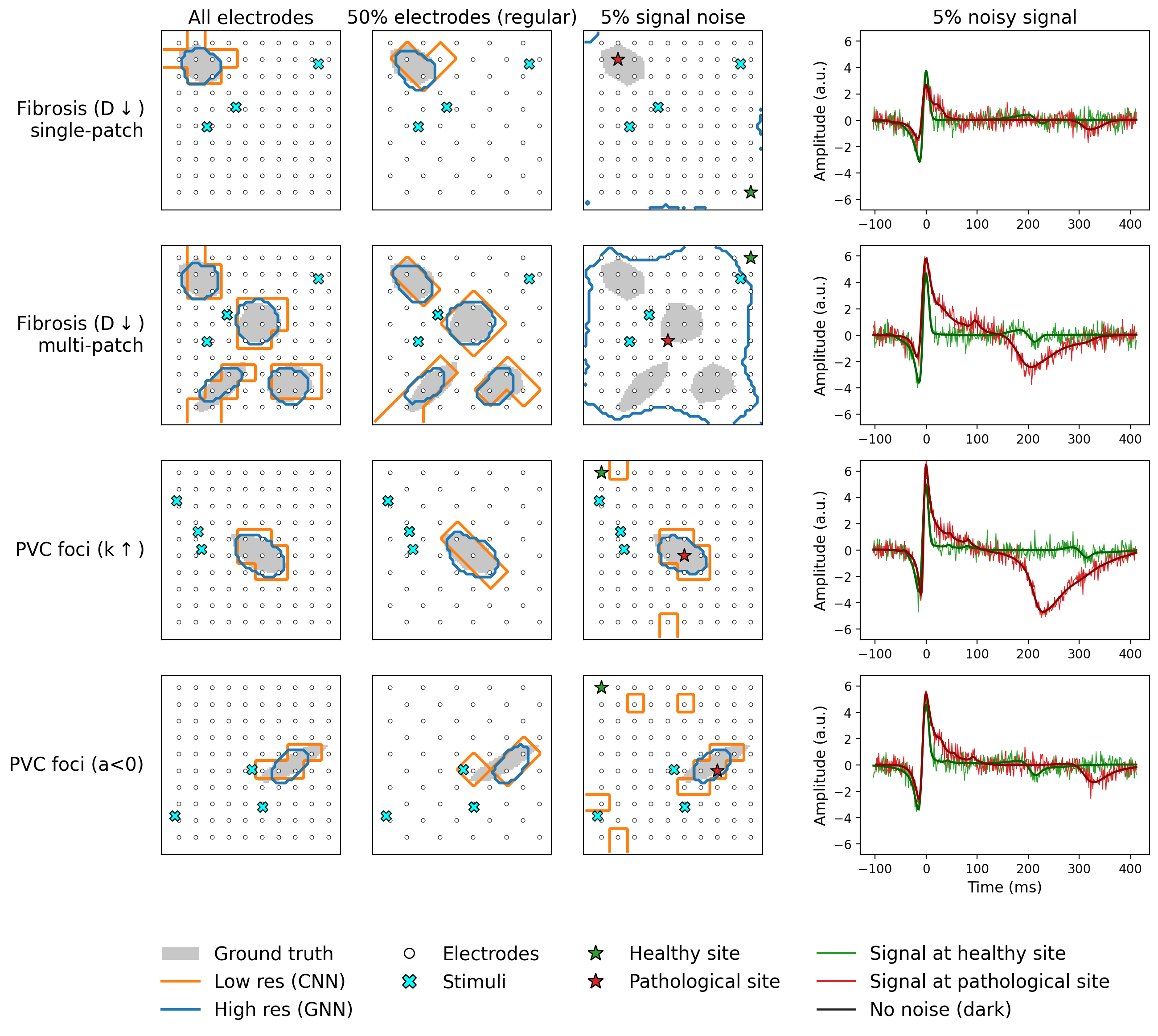}
    \caption{Visualisation of the low resolution model and our proposed GNN-based high resolution model for detecting single \& multi-patch lower $D$, higher $k$, and negative $a$. The performance with sparse electrodes and noisy signals are also shown. Examples of healthy signals and signals from the pathological regions (in the locations indicated by a matching colour star) with and without noise are shown in the rightmost column.}
    \label{fig:flat results}
\end{figure}

The performance of focal pathology detection across a range of metrics is summarised in Table~\ref{tab:results_metrics}, with example datasets displayed in Fig.~\ref{fig:flat results}. 

\begin{table}[htp!]
\centering
\caption{Performance evaluated by average precision, macro F1, precision/recall (of the abnormal class at a decision threshold of 0.5), and ROC AUC for low resolution (LR) and GNN-based high resolution (HR) model.}
\label{tab:results_metrics}
\begin{tabular*}{\textwidth}{@{\extracolsep{\fill}}lccccc}
\hline
 & \textbf{Average} & \textbf{Macro} & \textbf{Precision} & \textbf{Recall} & \textbf{ROC} \\
 & \textbf{precision} & \textbf{F1} & & & \textbf{AUC} \\
\hline
\multicolumn{6}{l}{\textbf{Low Resolution (LR) Model (CNN only)}} \\
\hline
\hspace{0.5cm}Fibrosis ($D\downarrow$), single patch & 0.76 & 0.80 & 0.45 & \textbf{0.96} & 0.990 \\
\hspace{0.5cm}Fibrosis ($D\downarrow$), multiple patches & 0.78 & 0.82 & 0.57 & \textbf{0.90} & 0.960 \\
\hspace{0.5cm}Rapid depolarisation ($k \uparrow$) & 0.74 & 0.86 & 0.61 & \textbf{0.92} & 0.991 \\
\hspace{0.5cm}High excitability ($a<0$) & 0.75 & 0.85 & 0.59 & \textbf{0.93} & 0.990 \\
\hline
\multicolumn{6}{l}{\textbf{High Resolution (HR) Model (GNN+CNN)}} \\
\hline
\hspace{0.5cm}Fibrosis ($D\downarrow$), single patch & \textbf{0.96} & \textbf{0.93} & \textbf{0.86} & 0.89 & \textbf{0.998} \\
\hspace{0.5cm}Fibrosis ($D\downarrow$), multiple patches & \textbf{0.93} & \textbf{0.91} & \textbf{0.84} & 0.84 & \textbf{0.986} \\
\hspace{0.5cm}Rapid depolarisation ($k \uparrow$) & \textbf{0.97} & \textbf{0.94} & \textbf{0.90} & 0.89 & \textbf{0.999} \\
\hspace{0.5cm}High excitability ($a<0$) & \textbf{0.95} & \textbf{0.93} & \textbf{0.89} & 0.84 & \textbf{0.998} \\
\hline
\end{tabular*}
\end{table}
The CNN-only LR model achieves good recall, indicating that it is able to pick up the pathological EGMs. However, the precision is consistently poor due to the limited spatial resolution of electrodes. In contrast, the GNN-based HR method localises the areas with abnormal tissue properties in detail beyond the electrode resolution, achieving higher F1 scores, demonstrating a good balance between precision and recall. It also shows better performance in threshold-independent metrics such as average precision\footnote{Here and throughout, average precision means precision averaged over ranked predictions as the threshold varies, weighted by changes in recall.} and ROC AUC across all pathology types. When detecting multiple fibrosis patches, both models perform slightly worse in the prevalence-independent metric, ROC AUC, compared to the single-patch cases, reflecting the greater difficulty of ranking several dispersed regions. 

\begin{figure}[ht!]
    \centering
    \includegraphics[width=1.0\linewidth]{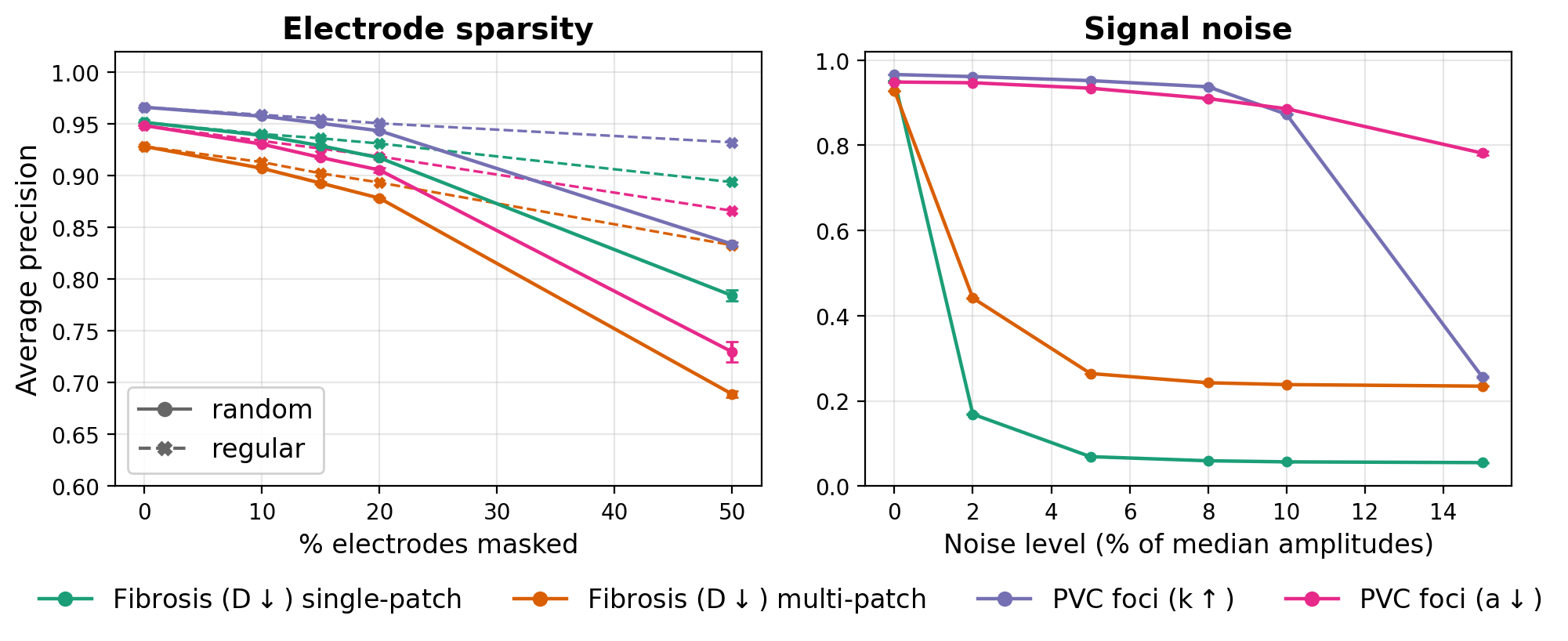}
    \caption{Average precision of the HR model with increasing electrode sparsity and signal noise. Performance for both random and regular downsampling of electrodes are shown. Error bars are computed over 3 different rounds of random sampling.}
    \label{fig:robustness metrics}
\end{figure}

The average precision of the HR model with respect to increasing electrode sparsity and signal noise is shown in Fig.~\ref{fig:robustness metrics}, with visual examples in Fig.~\ref{fig:flat results}. The HR model is generally robust against electrode sparsity across all three heterogeneity types, with average precision above 70\% up to half of the electrodes masked. Electrode sparsity reduces the model's ability to identify the gap between closely placed fibrotic patches in the multi-patch case. For noisy signals, the performance varies across pathologies, with fibrosis detection being particularly sensitive to noise for both LR and HR models. For high excitability and fast depolarisation detection, the LR model can be seen in Fig.~\ref{fig:flat results} to be less robust to noise, typically incorrectly confusing noise with small pathological regions. Quantitatively, LR model achieves a precision of 0.44 and 0.25, compared to the HR precision of 0.91 and 0.89 for higher $k$ and negative $a$, respectively, at 5\% noise.

\subsubsection{2D Curved Surfaces}

Table~\ref{tab:curved_surface} summarises the zero- (trained on the flat surfaces only) and few-shot performance of the HR model on curved surfaces for focal pathology detection. 
\begin{table}[htp!]
\centering
\caption{Zero-shot and few-shot performance on curved surfaces.}
\label{tab:curved_surface}
\begin{tabular*}{\textwidth}{@{\extracolsep{\fill}}lcccc}
\hline
\textbf{Metric} & \multicolumn{2}{c}{\textbf{Lower curvature}} & \multicolumn{2}{c}{\textbf{Higher curvature}} \\
& \multicolumn{2}{c}{(max $K$=$2.23\times10^{-3}~\text{mm}^{-2}$)} & \multicolumn{2}{c}{(max $K$=$1.19\times10^{-2}~\text{mm}^{-2}$)} \\
\cline{2-5}
& \textbf{Zero-shot} & \textbf{Few-shot} & \textbf{Zero-shot} & \textbf{Few-shot} \\
\hline
\multicolumn{5}{l}{\textbf{Fibrosis ($D\downarrow$)}} \\
\hline
\hspace{0.5cm}Average Precision & 0.857 & \textbf{0.986} & 0.632 & \textbf{0.958} \\
\hspace{0.5cm}ROC-AUC & 0.966 & \textbf{0.998} & 0.944 & \textbf{0.994} \\
\hspace{0.5cm}Max-F1 score & 0.822 & \textbf{0.937} & 0.704 & \textbf{0.880} \\
\hspace{0.5cm}Precision at Max-F1 & 0.785 & \textbf{0.966} & 0.693 & \textbf{0.971} \\
\hspace{0.5cm}Recall at Max-F1 & 0.862 & \textbf{0.911} & 0.715 & \textbf{0.805} \\
\hline
\multicolumn{5}{l}{\textbf{PVC foci ($k\uparrow$)}} \\
\hline
\hspace{0.5cm}Average Precision & 0.853 & \textbf{0.998} & 0.930 & \textbf{0.997} \\
\hspace{0.5cm}ROC-AUC & 0.966 & \textbf{1.000} & 0.989 & \textbf{1.000} \\
\hspace{0.5cm}Max-F1 score & 0.792 & \textbf{0.988} & 0.830 & \textbf{0.976} \\
\hspace{0.5cm}Precision at Max-F1 & \textbf{0.976} & \textbf{0.976} & 0.762 & \textbf{0.976} \\
\hspace{0.5cm}Recall at Max-F1 & 0.667 & \textbf{1.000} & 0.911 & \textbf{0.976} \\
\hline
\multicolumn{5}{l}{\textbf{PVC foci ($a<0$)}} \\
\hline
\hspace{0.5cm}Average Precision & 0.107 & \textbf{0.977} & 0.126 & \textbf{0.988} \\
\hspace{0.5cm}ROC-AUC & 0.508 & \textbf{0.996} & 0.598 & \textbf{0.998} \\
\hspace{0.5cm}Max-F1 score & 0.274 & \textbf{0.918} & 0.269 & \textbf{0.963} \\
\hspace{0.5cm}Precision at Max-F1 & 0.159 & \textbf{0.973} & 0.156 & \textbf{0.975} \\
\hspace{0.5cm}Recall at Max-F1 & \textbf{0.984} & 0.870 & \textbf{1.000} & 0.951 \\
\hline
\end{tabular*}
\end{table}
In the zero-shot setting, fibrosis (low $D$) and fast depolarisation (high $k$) detection both achieve average precisions of above 0.850, correctly identifying the approximate location although the shape is not accurate. Detection of heightened excitability (negative $a$) does not generalise well when directly applied to the curved surfaced in a zero-shot setting, at only 0.11 of average precision. 

\begin{figure}[htp!]
    \centering
    \includegraphics[width=0.96\linewidth]{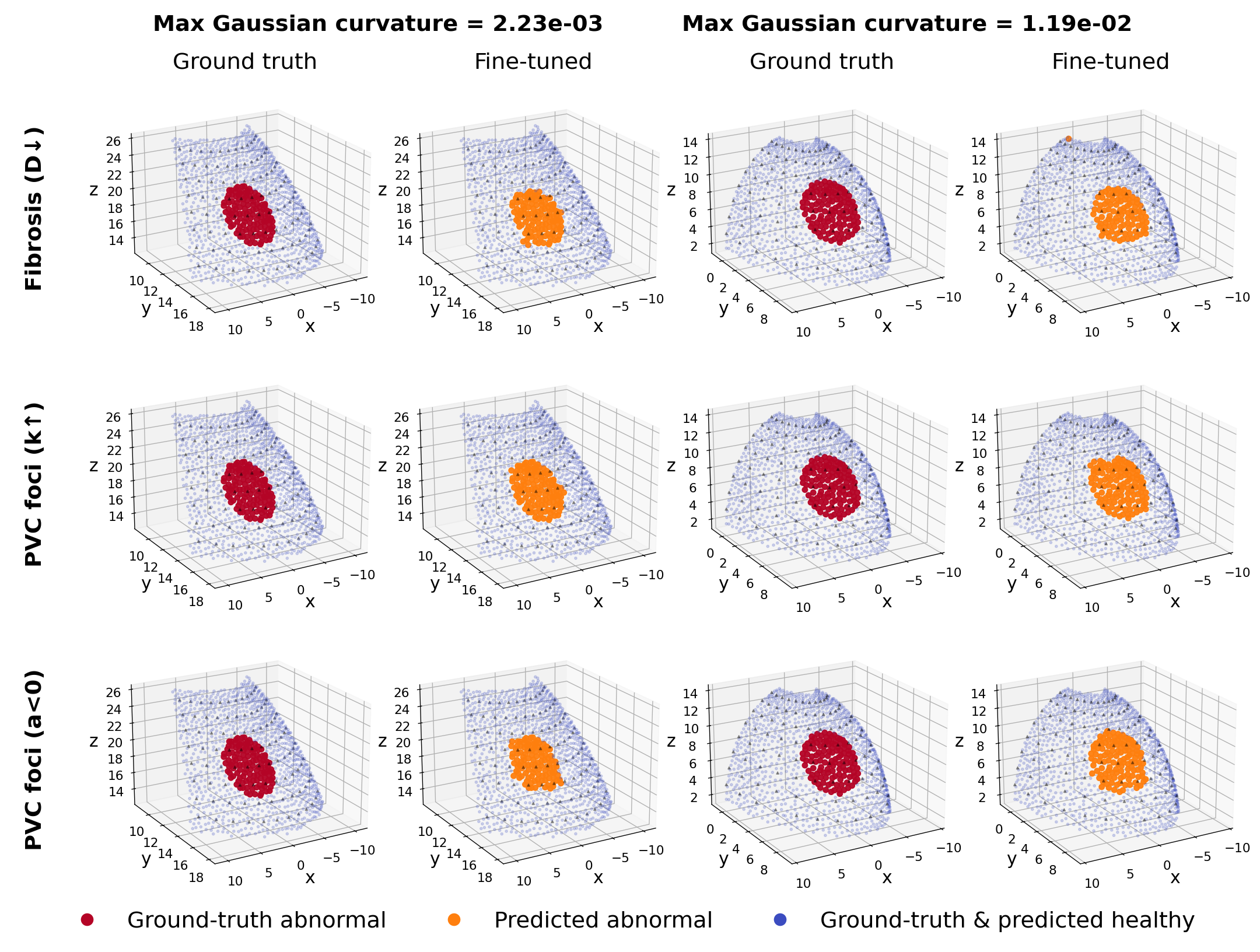}
    \caption{Visualisation of focal pathology detection results after few-shot (10 graphs) fine-tuning on curved surfaces.}
    \label{fig:curved}
\end{figure}

After fine-tuning just over 10 graphs, however, the model's performance improves and is exceptional in all three cases, achieving average precisions of above 0.97 across all three types of pathology and the two types of curved surfaces. The model has a higher performance in the lower curvature setting than in the higher curvature one. The few-shot performance visualised on two held-out graphs is shown in Fig.~\ref{fig:curved}, demonstrating the fine-tuned models' great performance.

\section{Discussion \& Conclusions}
We present a CNN and GNN-based method to characterise cardiac tissue properties from electrode measurements, specifically detecting regions of fibrosis, abnormal repolarisation, or heightened excitability, relevant for PVC ablation. 
The proposed method shows a good performance on simulated data, and demonstrates the potential to adapt to irregular distributions of sparse electrodes, and from flat to curved geometries with minimal fine-tuning, which is crucial for clinical settings and personalisation.

For flat surfaces, the LR 1D-CNN-only method does not see beyond the signals of one electrode at a time, whilst the HR CNN+GNN approach learns to process signals from all neighbouring electrodes, showing robustness against electrode sparsity. The GNN also naturally incorporates irregular layouts through the graph structure, which makes it more suitable for cardiac geometries compared to a 2D-CNN, which could use neighbouring electrode information but is better suited to regular grids. The HR method is more robust to noise compared to the LR one, as the latter often falsely picks up noisy signals as abnormalities. However, fibrosis detection is brittle against noise for both methods. It may be because the $D$ heterogeneities modelling fibrosis affect local conduction rather than single cell properties, making them harder to identify. 

We also show that the model trained on flat surfaces can be used on unseen curved surface directly or, for an improved performance, after minimal fine-tuning to learn artefacts from curvature. Generalising parameters learnt in planar tissue to curved geometries is non-trivial, because wave propagation speed and properties depend on tissue curvature~\cite{dierckx2011accurate}. In the zero-shot setting, the detection of heightened excitability ($a<0$) in curved surfaces is more difficult than the identification of the other two abnormal properties. This may be related to a dependency of excitability on the tissue curvature~\cite{dierckx2011accurate,connolly2018ventricular}.

Our GNN approach leverages the attention-based message passing layers, allowing each node to place learnt levels of importance on its neighbours as a function of the inverse of the distance between them, without explicitly fixing the attention weights. We hypothesise that this helps with the generalisability observed. These data suggest that the model learns features that transfer across geometries, relevant when we want to adapt to patient-specific geometries using limited data. Performance may be further increased if a variety of tissue curvatures are included in the training data.

Other techniques, beyond the GNN setup we used, could be employed to achieve higher-resolution classification performance, although they may not generalise as well to unseen geometries. These include implicit neural representations, which could parametrise the electrogram-derived features as a continuous function enabling data query at arbitrarily high spatial resolutions~\cite{sitzmann2020implicit}, classical spatial reconstruction methods such as radial basis function (RBF) interpolation, or physics-informed machine learning methods~\cite{chiu2024physics}. 

We note that, so far, both the planar and curved datasets are generated using the same isotropic Aliev–Panfilov model and a similar pathology-generation procedure. The current results therefore demonstrate transfer across synthetic geometries within one simulator. Further work will address how well the approach generalises to other electrophysiological models, electrophysiological heterogeneities, anisotropic conduction and geometries with non-zero thickness (as in the left ventricle). Ultimately, we expect to show that our model performs well in realistic patient anatomies, and in the presence of clinical EGM noise and artefacts. In the long run, it could play a role in the clinic to help identify PVC ablation targets using sinus rhythm data from PVC patients, improving the success rates and decreasing the duration of PVC ablation procedures.




\begin{credits}
\subsubsection{\ackname} We acknowledge computational resources and support provided by the Imperial College Research Computing Service (\url{http://doi.org/10.14469/hpc/2232}). 
We thank St George's Hospital Charity for research funding.
\end{credits}

\subsubsection{\discintname}
The authors have no competing interests to declare that are relevant to the content of this article. 
%
%
%
\bibliographystyle{splncs04}
\bibliography{ref}
%




\end{document}